\documentclass[aps,twocolumn,prl,preprintnumbers]{revtex4-1}
\usepackage{epsfig,amsmath}
\usepackage{subfigure}
\usepackage{graphicx}% Include figure files
\usepackage{dcolumn}% Align table columns on decimal point
\usepackage{stmaryrd}
\usepackage{mathrsfs}
\usepackage{pifont}
\usepackage{amsthm}
\usepackage{amssymb}
\usepackage{bm}
\usepackage{latexsym}
\usepackage{hyperref}
\usepackage{color}
\newcommand{\la}{\langle}
\newcommand{\ra}{\rangle}

\newcommand{\ga}{\gamma}

\newcommand{\si}{\sigma}

\newcommand{\om}{\omega}

\newcommand{\non}{\nonumber}
\newcommand{\pa}{\partial}

\def\pra#1{{ Phys.\ Rev. A\/} {\bf#1}}
\def\prb#1{{ Phys.\ Rev. B\/} {\bf#1}}
\def\pre#1{{ Phys.\ Rev. E\/} {\bf#1}}
\def\prl#1{{ Phys.\ Rev.\ Lett.} {\bf#1}}
\def\pr#1{{ Phys.\ Rev.} {\bf#1}}

\def\pla#1{{ Phys.\ Lett. A\/} {\bf#1}}
\def\rmp#1{{ Rev. \ Mod. \ Phys.} {\bf#1}}

\begin{document}

\title{Control of decoherence with no control}

\author{Jun Jing$^{1,2}$ \footnote{Email address: jingjun@shu.edu.cn} and Lian-Ao Wu$^{2,3}$ \footnote{Email address: lianao\_wu@ehu.es}}

\affiliation{$^{1}$Institute of Theoretical Physics and Department of Physics, Shanghai University, Shanghai 200444, China\\ $^{2}$Department of Theoretical Physics and History of Science, The Basque Country University (EHU/UPV), PO Box 644, 48080 Bilbao\\ $^3$ Ikerbasque, Basque Foundation for Science, 48011 Bilbao Spain}

%\date{\today}

\begin{abstract}
%Throughout history, 
Common philosophy in control theory is the control of disorder by order. It is not exceptional for strategies suppressing quantum decoherence. Here we predict an anomalous quantum phenomenon. Suppression of decoherence can be made via more disordered white noise field, in particular white Poissonian noise. The phenomenon seems to be another anomaly in quantum mechanics, and may offer a new strategy in quantum control practices.
\end{abstract}

\maketitle

Decoherence is the deterioration of quantum information in a system due to inevitable interactions with the environment or bath \cite{Breuer,Leggett,Preskill,Gardiner,Zurek}. Suppression of decoherence is one of the paramount challenges in quantum control practices and requires accurate control of the system dynamics by time-dependent fields such as laser pulse sequences. Here we predict an anomaly: suppression of decoherence can be made via uncontrollable white noise field. By increasing the strength of noise signals, a two-level system becomes less coupled to its environment and even remains nearly intact for a period of time. The aberrant effect reveals a different physical mechanism in quantum control theory, and in practice may offer the possibility of control by uncontrollable white noise: control of decoherence with no control.

Noise is a source of disorder. White noise, whose spectrum has equal power within any equal interval of frequencies, is the extreme of disorder in comparison with coloured noise. Over a decade ago, people began to notice in classical systems that noise leads not only to nuisance but also to advantages. A remarkable example is that an external coloured noise can suppress the intrinsic system white noise \cite{Rubi,Walton}. While it is a surprise, the phenomenon fits well with common philosophy: control of disorder by order. The philosophy has been carried out in classical noise control and extended to suppression of decoherence in quantum dynamical processes. External field control of quantum decoherence dates back to the spin echo technique \cite{Hahn}. This technique is used to suppress the inhomogeneous spin dephasing by applying a $\pi$ inversion pulse and has been developed to tackle general decoherence \cite{Wiseman,Kofman,Viola,Viola1,Wu4,Uhrig,Lidar4,Zhang,PQ}.  The philosophy, control of disorder by order, remains the same for either the classical or the quantum mechanical. Now the predicted anomaly is opposite to the common philosophy. Our discovery reveals that the most disordered white noise can control less disordered decoherence, which can be characterized by a quantum stochastic process with colored noise. Seeing that the setting in use is exclusively quantum mechanical, it appears that the phenomenon is another anomaly in quantum mechanics, in particular in open quantum system.

In quantum mechanics, a dynamical process of the system plus the environment is governed by the total Hamiltonian,
\begin{equation}
H_{\rm tot}=H_{\rm S}(t)+H_{\rm B}+H_{\rm SB},
\end{equation}
where $H_{\rm S}(t)$ and $H_{\rm B}$ are the system Hamiltonian, embedded with white noise, and the environment Hamiltonians. The system dynamics is normally characterized by a variety of master equations, after tracing out the environment. The system-bath interaction $H_{\rm SB}$ is the source of decoherence.

\bigskip
\noindent\textbf{\large {Results}}\\
\noindent\textbf{Protocol of decoherence suppressions with no control.}
\begin{figure}[htbp]
\centering
\includegraphics[width=3.5in]{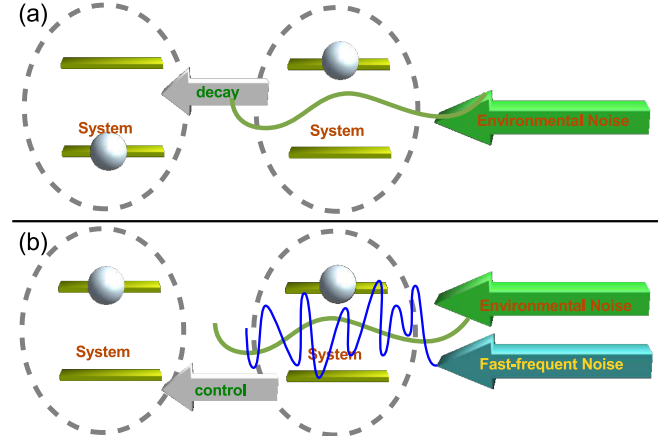}
\caption{Sketches of (a) dissipative process of a two-level system induced by its environment, (b) dissipation suppressed by white noise.}\label{dia}
\end{figure}
We now come to introduce our protocol of decoherence suppressions, in particular suppression of dissipation. Dissipation is a decoherence process where the populations of quantum states vary due to involvement of the environment. Provided that a two-level system is in its excitation state, the environment induces the system to give off energy and to decohere to the ground state [Fig.\ref{dia}(a)]. If nature happens to have the desired white noise or we artificially apply such noise signals on the system, the decoherence can be suppressed [Fig.\ref{dia}(b)]. This required noise is described by $c(t)=\eta(J, W, t)$, where $\eta$ is white noise, in particular the biased Poissonian white noise and $J$ is the noise strength \cite{Hanggi}. We name the average time interval between two neighbour noise signals as $1/W\equiv T/n$, where $T$ is a time scale, different for different systems, and $n$ is the noise arrival number \cite{Hanggi}. When $1/W$ goes to zero, $c(t)$ corresponds to the continuous-time white noise process, where $J$ is the only parameter. If $1/W$ is finite, $c(t)$ can be the biased Poissonian white shot noise, which is essentially different from well-controlled pulse sequences, idealized or non-idealized. We can tune the parameter $W$ towards the continuous-time limit. In what follows, we shall incorporate $c(t)$ into the two-level system to display physical effects.

\bigskip
\noindent \textbf{Fidelity preserved by white noise.}
Consider a dissipative model for the two-level system, described by a non-Hermitian Hamiltonian in the exact quantum Stochastic Schr\"odinger equation [See \textbf {Method}] \cite{Diosi1,Diosi2},
\begin{equation}\label{Heff}
H_{\rm ss}(t)=[\omega+c(t)]\si_z/2+iz_t^*g\si_--igQ\si_+\si_-,
\end{equation}
where $\om$ is the bare-energy spacing and $g$ is the coupling strength between system and environment. $c(t)$ is the above-mentioned white noise signal and $Q$ satisfies a nonlinear differential equation $\dot{Q}(t)=g\ga/2+[-\ga+i\om+ic(t)]Q+gQ^2$, with a boundary condition $Q(0)=0$ \cite{PQ}.  The correction function of this process is $G(t,s)=\frac{\ga}{2}e^{-\ga|t-s|}$. Here $\ga$ characterizes the environmental memory in the Ornstein-Uhlenbek process and is inversely proportional to the environmental memory time. The values of $\ga$ can be used to somehow determines the degree of Markovianity. The larger $\ga$ is, the more Markovian the environment would be. $\ga \rightarrow \infty $ corresponds to the environmental white noise model and indicates the Markov limit. The environmental noise is formulated as $z^*_{t+\Delta t}=z^*_t-\ga z^*_t\Delta t+\sqrt{\Delta t/2}\ga w^*$, where $w^*$ is a complex Wiener process \cite{PQ}.

%satisfying $M[w^*]=0$ and $M[w_tw^*_s]=\delta(t,s)$.

Suppose that the system initial state is $|\psi_0\ra=|1\ra$. The fidelity $\mathcal{F}(t)\equiv\sqrt{\la\psi_0|\rho_t|\psi_0\ra}$, qualifying the survival probability, evolves according to
\begin{equation}\label{qubit}
\mathcal{F}(t)=e^{-\int_0^tds\mathcal{R}[Q(s)]},
\end{equation}
where $\mathcal{R}[\cdot]$ is the real part of the input function. Below we will numerically study behaviors of the fidelity in the short-time regime to determine the effect of white noise signals.

\begin{figure}[htbp]
\centering
\subfigure{\label{dnM}
\includegraphics[width=3in]{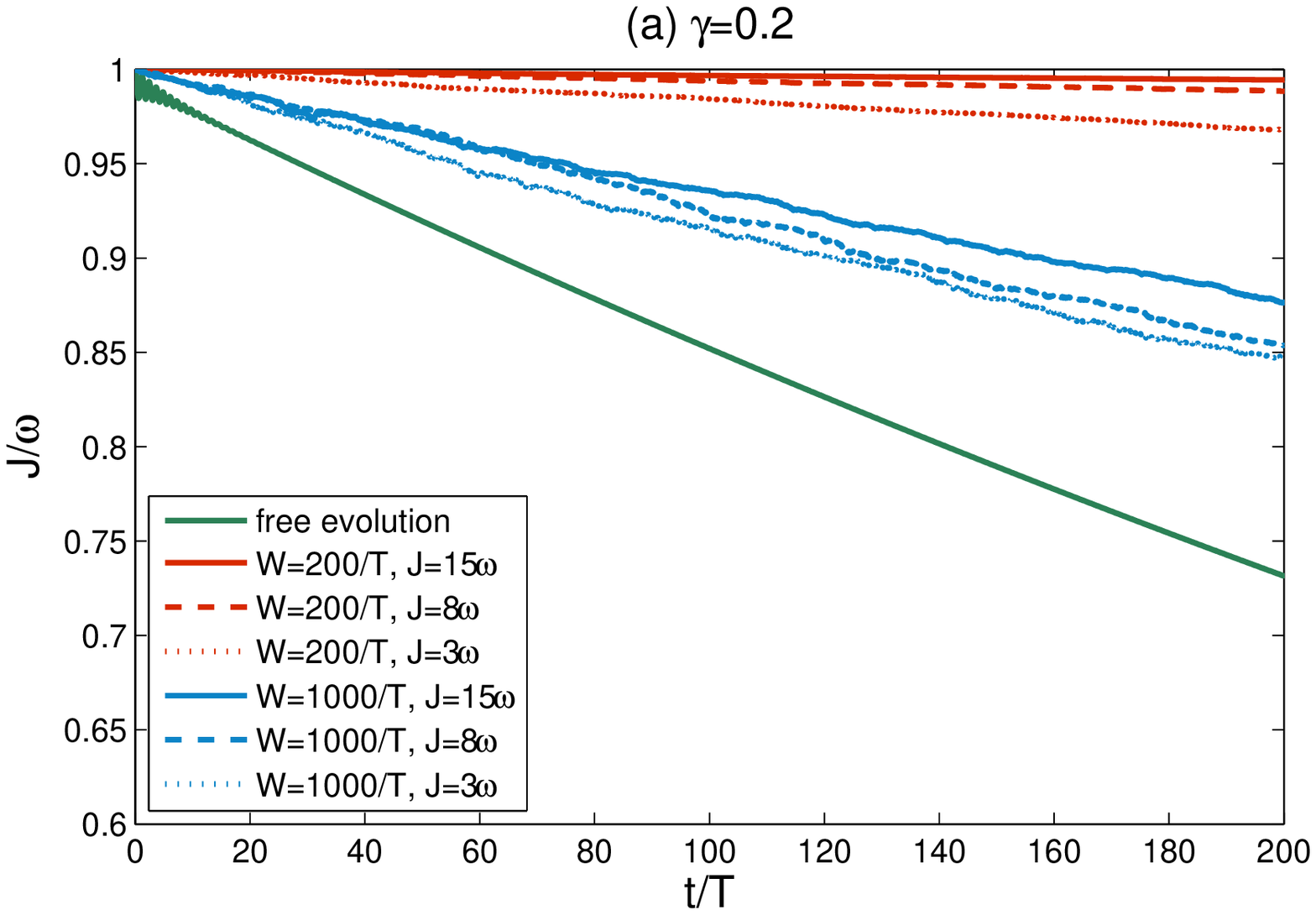}}
\subfigure{\label{dM}
\includegraphics[width=3in]{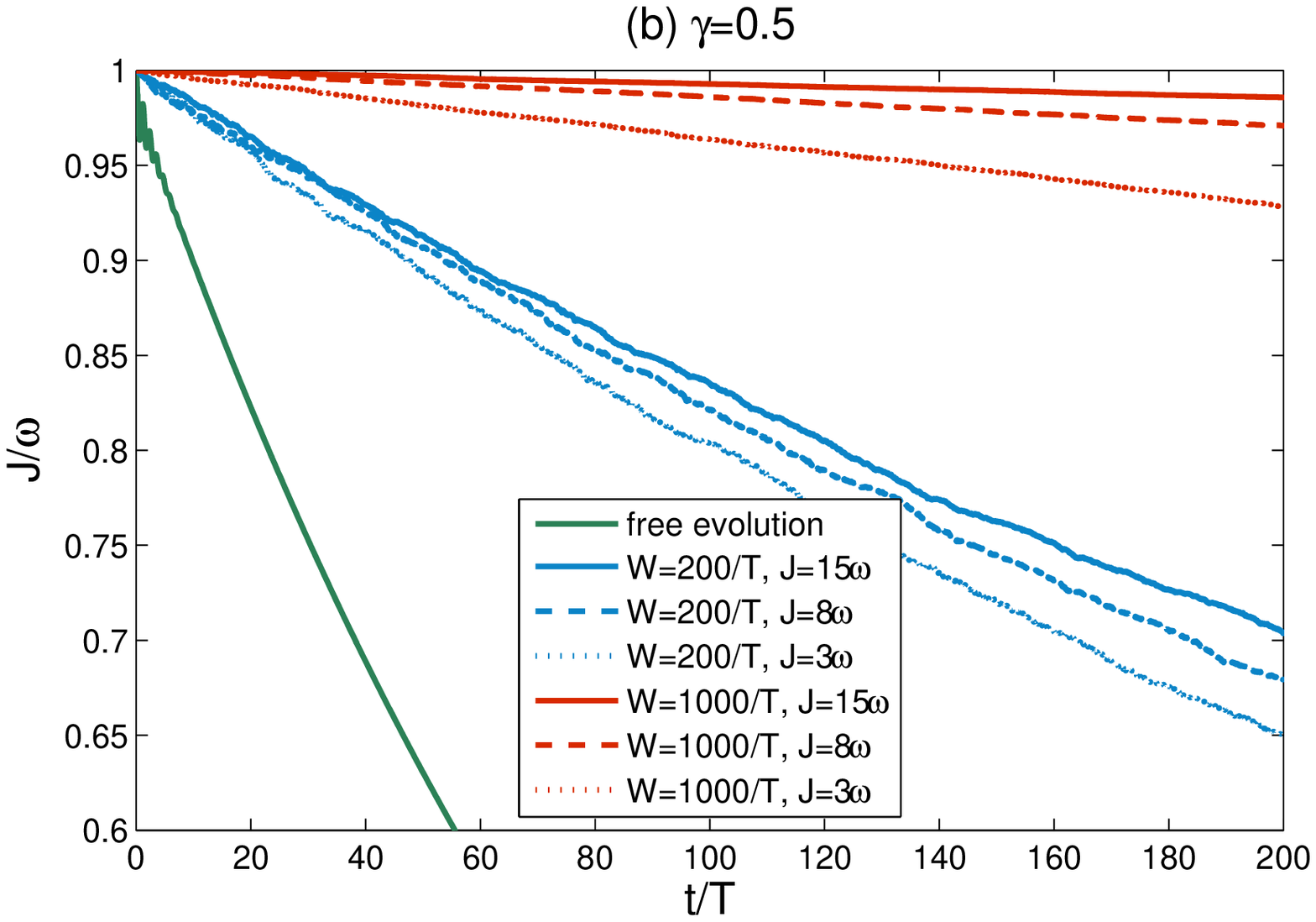}}
\caption{$\mathcal{F}(t)$ vs time. The environmental memory parameter $\ga$ is taken as $0.2$ and $0.5$ respectively. We compare the free dynamics (green-solid lines) with those subject to different noises, red lines $W=1000/T$ and blue lines $W=200/T$. The solid lines, dashed lines and dotted lines represent different strength $J=15\omega$, $8\omega$ and $J=3\omega$ respectively. Here $\omega T=5$ and $g=0.4\omega$.}\label{FyD}
\end{figure}

Figures \ref{dnM} and \ref{dM} show $\mathcal{F}(t)$ vs time for $\ga=0.2$ and $0.5$, subject to different $c(t)$. Suppression of dissipation is excellent in all cases with a bigger value of $W$, yet better in less Markovian environment. On contrast, the fidelity decays rapidly with time in absence of $c(t)$. Different values of $W$ match different physics. Smaller $W$ corresponds to a sequence of shots with random amplitudes and sparser random arrival moments, illustrated by $W=200/T$. It suppresses dissipation to some extent, but not as efficient as big $W$. The results show that for different values of big $W$ (here $W=1000/T$) the fidelity is asymptotic to an identical curve, as shown by the red solid lines in both figures. These red soild curves corresponds to the continuous-time white noise process. It shows that continuous-time white noise signals prevent the system from dissipation.

The quantity of suppression by continuous-time white noise relies mostly on the noise strength $J$ and environmental non-Markovianity $\ga$. The two figures show clearly that the larger $J$ is, the better the quality of suppression is. %The control efficiency is not very sensitive to the strength $J$, although a stronger suppression could be achieved by a bigger $J$.
To this end, we may employ $J$ as large as physically possible. Even so, the additional average energy spacing of the stochastic $c(t)$, determined by $J$. remains small. The parameter $\ga$ seriously influences the quality of suppression as well. Suppression works worse at $\ga=0.5$ than at $\ga=0.2$, and even worse when the environment becomes more Markovian. Eventually, the suppression will not work in a complete Markovian environment ($\ga\rightarrow\infty$) corresponding to quantum white noise. Interestingly, this shows that white noise cannot suppress white noise, i.e. complete disorder cannot suppress complete disorder.

\begin{figure}[htbp]
\centering
\subfigure{\label{TR5}
\includegraphics[width=3in]{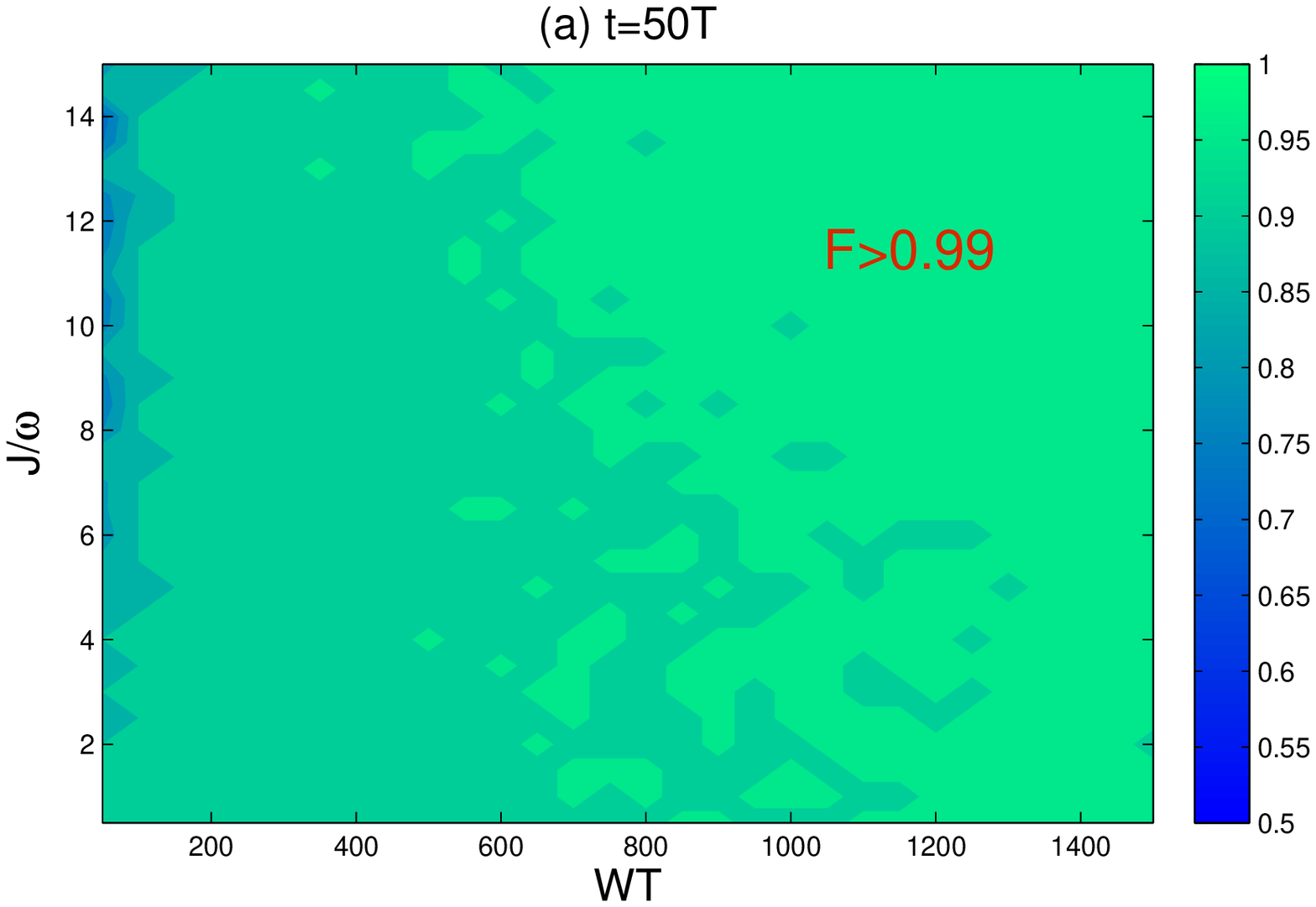}}
\subfigure{\label{TR10}
\includegraphics[width=3in]{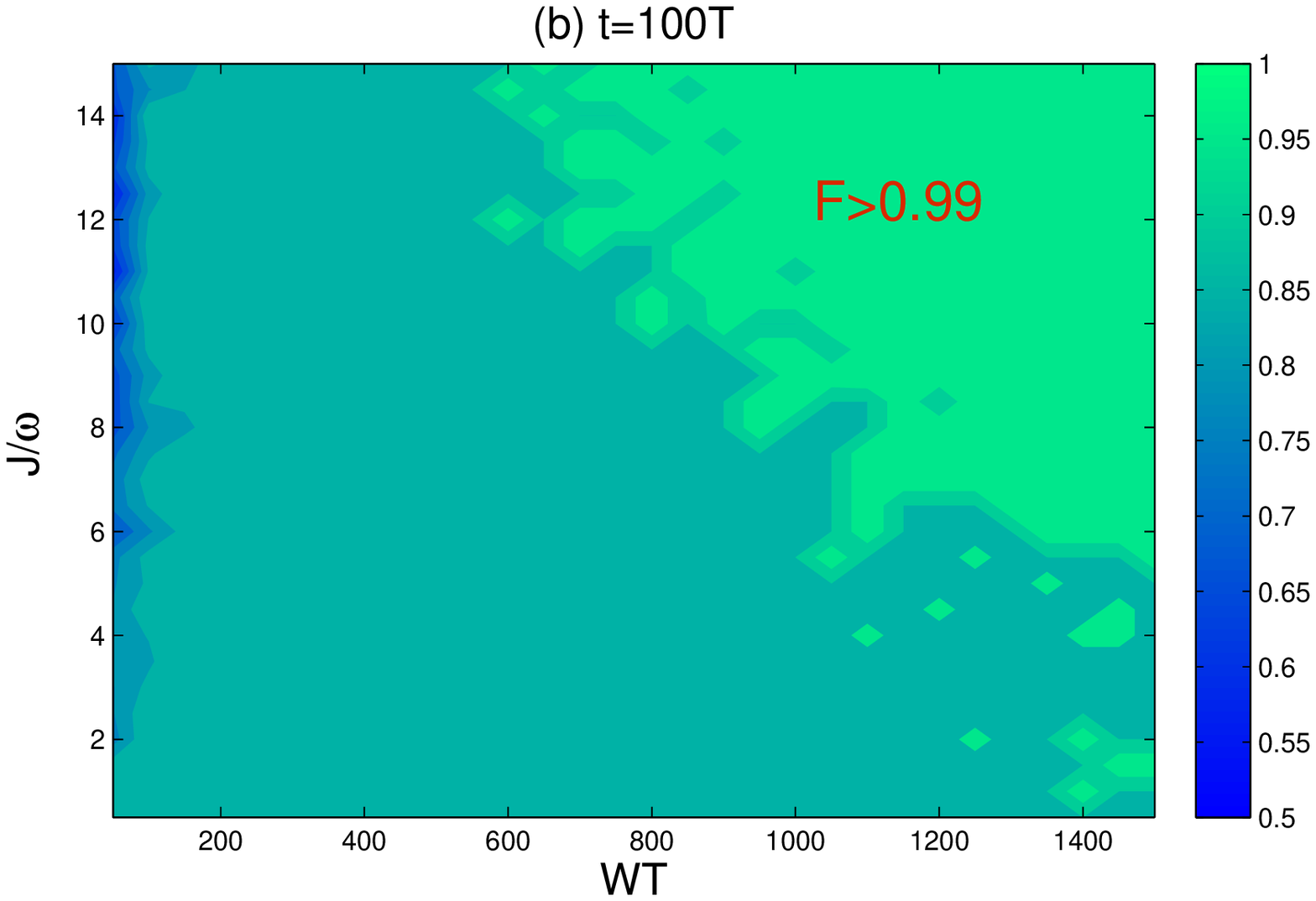}}
\caption{Parameter thresholds and effective regions at two moments. Fidelity where $\mathcal{F}>0.99$ is distinguished. The non-Markovian environment is characterized by $\ga=0.2$. Here $\omega T=5$ and $g=0.4\omega$. }\label{TR3d}
\end{figure}

%\begin{table}[h!]
%\begin{tabular}{|c|c|c|c|c|c|} \hline
%& Rydberg atom & molecule & Electron spin & Nuclear spin & Superconducting qubit \\ \hline
%$\omega$ & $10^9Hz$ & $10^{12}-10^{15}Hz$ & $10^9-10^{10}Hz$ & $10^6-10^7Hz$ & $10^9-10^{10}Hz$ \\ \hline
%$1/\omega$ & $10^6fs$ & $1-10^3fs$ & $10^5-10^6fs$ & $10^8-10^9 fs$ & $10^5-10^6fs$ \\ \hline
%\end{tabular}
%\caption{Characteristic frequencies of different effective two-level systems} \label{qutab}
%\end{table}

Now we look into the detailed roles that the noise strength $J$ and $W$ play. Figure \ref{TR3d} plots the fidelity contour as a function of $J$ and $W$ at two time moments $t=50T$ and $t=100T$. The regions where $\mathcal{F}>0.99$ are highlighted. The fidelity is saturated at $W>600/T$ for both figures, where a discrete random pulse sequence becomes the continuous-time white noise. While the fidelity is excellent even when $J$ is not very big, it seems not to be saturated with $J$. The bigger $J$ is, the better the fidelity is.

The numerical results presented by figures could be valid for various physical systems with corresponding characteristic values of $T$. For example, for a superconducting flux qubit $\om$ is approximately $10^9-10^{10}$Hz \cite{Xiang}. The relaxation time is $T_1\approx 1 \mu s $ so that the time scale $T\approx 5 nm$ and the dimensionless $\om T \approx 5$ as taken in Fig. \ref{dnM}. The required noise strength $J$ should be more than $10^9$Hz in order to successfully suppress decoherence.

\bigskip
\noindent\textbf{Discussion.}
The perfect suppression could be justified by the following argument. By the integrating over the equation of motion (\ref{qsd}), we will always be able to write,
\begin{equation}\label{ana}
\la\psi_0|\psi_\infty\ra-1=\sum_{\alpha\beta}\int_0^\infty N_{\alpha\beta}(t)h_{\alpha\beta}(t)dt,
\end{equation}
for a long time limit, where $\alpha$'s denote a set of complete bases. Here $N(t)$ and $h(t)$ represent a noisy matrix and a system dynamical matrix (see example in {\bf{Method}}). If each nonzero element of $N$ is a fast-varying noise function of time whereas $h$ is much slower and weaker, the integral of each term $N_{\alpha\beta}(t)h_{\alpha\beta}(t)$ could be zero since strong oscillation of the fast one will wash out accumulation of the slow function. Our numerical results show that $N_{\alpha\beta}(t)$ is much stronger than $h_{\alpha\beta}(t)$ when the noise strength $ J$ is strong. In our case, $N(t)\propto e^{-i\int_0^t c(s) ds}$ and $h(t)=- Q(t)\la\psi_0|\psi_t\ra $ are c-numbers. The bigger values of $J$, $h(t)$ is slower and weaker than $N(t)$. When $h(t)$ is so slow in comparison with $N(t)$ that it can be treated as time independent, it is easily to prove $\int_0^\infty N(t) dt =0 $ for white noise $c(t)$, which appears to be a unique feature of continuous-time white noise.

The signal $c(t)$ mimics white noise in natural processes and is also associated with the natural dephasing processes. Our results show that the existence of $c(t)$, or dephasing, significantly inhibits the dissipation. This well explains the reality where the relaxation time $T_1$ is longer than the dephasing time $T_2$ for all systems.

\bigskip

\noindent\textbf{\large{Method}} \\
Consider the system-bath interaction $H_{\rm SB}=LB^\dag+L^\dag B$ for simplicity. The exact stochastic Schr\"odinger equation is \cite{Diosi1,Diosi2,Jing}:
\begin{equation}\label{qsd}
i\pa_t\psi_t(z^*)=H_{\rm ss}(t)\psi_t(z^*).
\end{equation}
where $H_{\rm ss}=H_S (t)+iLz_t^*-iL^\dag\bar{O}$ is an exact system Hamiltonian, for instance $L=g\sigma_{-}$ for our two-level system. $\bar{O}$ is a combination of system operators and environmental noises satisfying consistency conditions \cite{Diosi1}. The operator describes the environmental influence without invoking the master equation. Every quantum trajectory is accompanied by a special process $z^*$, thus the system density matrix is recovered by $\rho_t=M[|\psi_t(z^*)\ra\la\psi_t(z^*)|]$.

%The suppression mechanism could be understood by the following brief analysis.
Suppose that the system state is initially at $|\psi_0 \ra=|\mu\ra$, one basis of a completed set $|\nu\ra$'s. The system Hamiltonian and the coupling operator could be generally expressed by $H_{\rm S}(t) =\sum_{\nu}\om_\nu(t)|\nu\ra\la\nu|$ and $L=\sum_{\mu\ne\nu}C_{\mu\nu}|\mu\ra\la\nu|$ respectively,
\begin{equation}\label{ana1}
h_{ \alpha \beta} =z_{ t }^{ * }C_{ \beta \alpha} \delta _{ \beta \mu} \langle \alpha |\psi _{ t }\rangle -\sum _{ \mu '\nu ' } C_{ \alpha \beta^*}\delta _{ \alpha \mu}\delta _{ \beta \nu ' }\langle \nu '|\bar { O } |\mu '\rangle \langle \mu '|\psi _{ t }\rangle
\end{equation}
and $N_{\alpha \beta}\equiv  e^{-i\int_0^t[\om_\alpha(s) -\om_\beta(s)]ds}$. The white noise is embedded in the difference $\om_\alpha-\om_\beta$. $\mathcal{F}=\sqrt{\la\psi_0|M[|\psi_t\ra\la\psi_t|]|\psi_0\ra]}$ is obtained by an ensemble average over the integral results in Eq.~(\ref{ana}). For the two-level system initially at $|1\ra$, we can specifically write (\ref{ana}) as,
\begin{equation}
\la\psi_0|\psi_\infty\ra-1=-\int_0^\infty Q(t)N^*\la\psi_0|\psi_t\ra dt
\end{equation}
where $N(t)=e^{-i\int_0^t[\om+c(s)]ds}$.

We employ the biased Poissonian white noise \cite{Hanggi,Hanggi2}, $c(t)=\sum_jx_j\delta(t-t_j)$ satisfying the following statistical properties:
\begin{eqnarray}
&&M[c(t)]=JW, J=M[x_j]; \\  \non
 &&M[c(t)c(s)]-M[c(t)]M[c(s)]=WM[x_j^2]\delta(t-s),
\end{eqnarray}
where $x_j$'s are noise heights. Details of numerical realization of the noise can be found in Ref. \cite{Hanggi2}.

\noindent\textbf{Acknowledgements}\\
We acknowledge grant support from the National Natural Science Foundation of China under Grant No. 11175110, the Basque Government (grant IT472-10) and the Spanish MICINN (Project No.  No. FIS2012-36673-C03- 03).

\noindent\textbf{Author contributions}\\
J.J. contributed to numerical and physical analysis and prepared the first version of the manuscript and L. -A. W. to the conception and design of this work. Both authors wrote the manuscript.

\noindent\textbf{Additional Information} \\
Competing financial interests: The authors declare no competing financial interests.

\end{document}